\documentclass[twocolumn,fleqn,10pt]{wlscirep}
\usepackage[utf8]{inputenc}
\usepackage[T1]{fontenc}
\usepackage{amssymb,amsmath,gensymb}
\usepackage{lipsum}
\usepackage{CJKutf8}
\usepackage{jabbrv}
\usepackage[normalem]{ulem}
\usepackage{changes}
\usepackage{chngcntr}
\usepackage{epsfig}
\usepackage{amssymb}
\usepackage{soul}
\usepackage{color}
\usepackage{float}
\usepackage{textcomp}
\usepackage{amstext}
\usepackage{graphicx}
\usepackage{gensymb}
\usepackage{graphics}\usepackage{subfigure}\usepackage{longtable}\usepackage{pstricks}\usepackage{dcolumn}\usepackage{bm}
\usepackage{siunitx}
\usepackage{wasysym}
\usepackage[font=small,labelfont=bf]{caption}

\usepackage{booktabs}  
\usepackage{xcolor}

\makeatletter
\newcommand*{\balancecolsandclearpage}{%
  \close@column@grid
  \cleardoublepage
  \twocolumngrid
}
\makeatother

\newcommand{\LNOthree}{La$_4$Ni$_3$O$_{10}$}
\newcommand{\LNO}{La$_5$Ni$_4$O$_{13}$}
\newcommand{\LNOr}{La$_5$Ni$_4$O$_{10}$}
\newcommand{\LNOfive}{La$_6$Ni$_5$O$_{16}$}
\newcommand{\LNOfiver}{La$_6$Ni$_5$O$_{12}$}

\title{Ambient-pressure superconductivity in bulk single crystals of \LNOr}

\author[1]{R.~Zhang} 
\author[1]{P.~Reiss}
\author[1]{A.~Krajewska}
\author[1]{M.~Isobe}
\author[1]{H.~Takagi}
\author[1,*]{P.~Puphal}
\affil[1]{Max-Planck-Institute for Solid State Research, Heisenbergstrasse 1, 70569 Stuttgart, Germany}
\affil[*]{puphal@fkf.mpg.de}

\begin{abstract}
Superconductivity in the Ruddlesden–Popper (RP) nickelates has so far been confined either to multilayer compounds under high pressure or to square-planar infinite-layer-type films stabilized by epitaxial strain. Here, we report the high-pressure high-temperature (HPHT) growth of bulk single crystals of the quadruple-layer RP nickelate La$_5$Ni$_4$O$_{13}$ ($n=4$) and reproduce the quintuple-layer La$_6$Ni$_5$O$_{16}$ ($n=5$) synthesis, phases previously accessible only as thin films. Topotactic reduction of La$_5$Ni$_4$O$_{13}$ single crystals with CaH$_2$ yields the square-planar La$_5$Ni$_4$O$_{10}$, which we show to be a bulk, ambient-pressure superconductor with $T_\mathrm{c}^\mathrm{onset}\approx18.7$\,K, a shielding fraction above 60\%, and a specific-heat jump that confirms a bulk pairing transition. In transport, all crystals show an anomaly near 90~K reminiscent of a density-wave like transition, which may arise from a partially gapped Fermi surface or from the non-superconducting volume fraction. The upper critical field shows a small anisotropy, $\gamma \approx 2.4$, which remains finite as $T \to T_c$, indicating three-dimensional superconductivity with substantial coupling between the four-layer blocks. These results identify the square-planar nickelates $R_{n+1}$Ni$_n$O$_{2n+2}$ as a bulk platform in which the layer number alone tunes the nickel valence into the superconducting regime, without chemical substitution, pressure, or epitaxial constraint.
\end{abstract}
\begin{document}

\flushbottom
\maketitle

\thispagestyle{empty}

The RP nickelates La$_{n+1}$Ni$_n$O$_{3n+1}$ provide a model system for studying how dimensionality and band filling jointly control the ground state of a correlated transition-metal oxide. As the number of perovskite layers $n$ increases, the nominal Ni valence rises from $2+$ at $n=1$ towards $3+$ as $n\rightarrow\infty$, following $3-1/n$. This structural progression generates a rich phase diagram: La$_2$NiO$_4$ ($n=1$) is an antiferromagnetic insulator, the $n=2$ and $n=3$ compounds are metals that host density-wave transitions~\cite{khasanov2025pressure,zhang2017large} and become superconducting under pressure~\cite{Sun2023,Puphal2024,Wang2024,Zhu2024,Shi2025}, and LaNiO$_3$ ($n\rightarrow\infty$) is a paramagnetic metal at ambient conditions. The series therefore spans, within a single chemical family, the full range of behaviour from Mott-like insulator to correlated metal to conventional metal, making it an unusually direct testbed for theories that link interlayer coupling, orbital occupation, and superconductivity. Even hybrid forms of RP nickelates exist as single crystals with a mixed stacking of $n$ with alternating monolayer and bilayer, i.e. 1212 \cite{Li2024a,Shi2025}, and alternating monolayer and trilayer 1313 \cite{Puphal2024,Chen2024,Huang2026, Wang2024e} exist. They exhibit Ni charge mixtures and unique density wave transitions \cite{Khasanov2025b,zhang2026spin_density_wave}, yet under pressure the T$_C$ is reduced from the monolayer blocks but superconductivity remains \cite{Shi2025,Huang2026}.

\begin{figure*}
    \centering
    \includegraphics[width=1\linewidth]{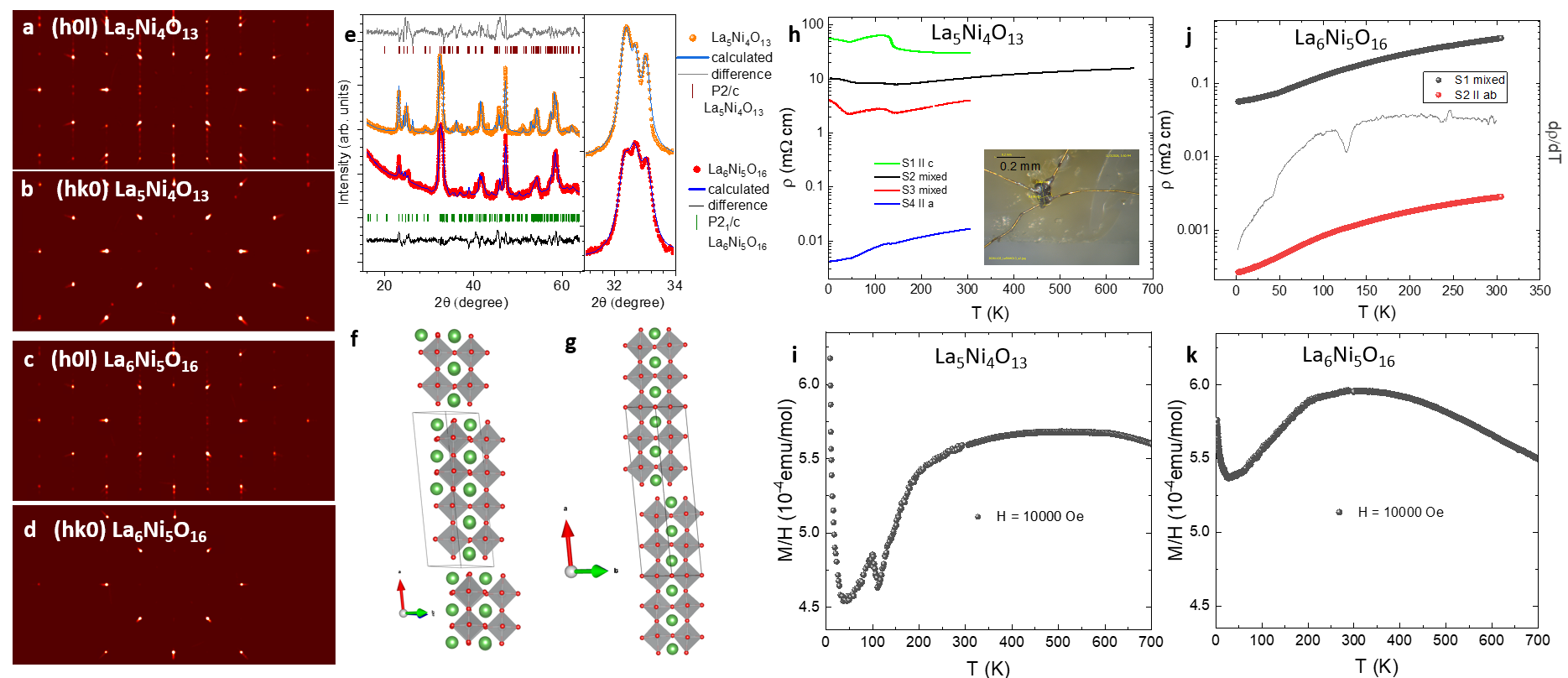}
    \caption{\textbf{Synthesis of bulk high order RP nickelates. a,b}
    Single-crystal XRD maps of the (h0l) and (hk0) planes of \LNO{}. and \textbf{c,d,} of \LNOfive{}. \textbf{e,} Comparative PXRD patterns of crushed single crystals of La$_5$Ni$_4$O$_{13}$ and La$_6$Ni$_5$O$_{16}$ with underlying Rietveld refinements; the right panel shows an enlarged view of a selected region showing the difference via the (114) reflex distance to the (020) reflex. \textbf{f,g,} Resulting structure of \LNO{} and \LNOfive{}. \textbf{h,i,} Temperature dependent resistance and magnetic susceptibility in an applied field of 10000 Oe of \LNO{} and \textbf{j,k,} of \LNOfive{}.}
    \label{RP}
\end{figure*}

In bulk, however, this ladder of compounds stops at $n=3$. Members with $n\geq4$ are not thermodynamically stable under conventional high temperature ambient-pressure synthesis. Essentially everything known about the higher-$n$ members comes from metastable epitaxial films~\cite{Li2020APL}. For La$_5$Ni$_4$O$_{13}$ in thin-film form, even basic structural parameters, the stacking sequence of the four layers, the number of crystallographically distinct Ni sites, and the oxygen content, must be inferred rather than directly measured, with substrate-imposed strain always present as a confounding variable. A bulk single crystal would resolve these ambiguities. The $n=5$ member was only very recently obtained in bulk form under high pressure~\cite{Shi2026}; the $n=4$ member, the subject of this work, has not previously been reported in bulk.

There is a second motivation for growing these crystals. Topotactic removal of the apical oxygen converts the RP phases into square-planar compounds with $n>1$: $R_{n+1}$Ni$_n$O$_{2n+2}$, in which the nominal Ni valence, $1+1/n$, is again set purely by the layer number, falling from $+1.5$ at $n=2$ toward $+1$ as $n\rightarrow\infty$. In thin films, superconductivity has been reported at $n=5$~\cite{Pan2022}, with signatures extending from $n=4$ to $n=8$~\cite{Pan2026}: the range over which the Ni valence falls within the putative superconducting dome observed in infinite-layer (IL) nickelates \cite{Zeng2020, Puphal2025}. Thus, the layer number offers an alternative, purely structural handle: it tunes the Ni valence by self-doping, without any need for aliovalent substitution.  Reproducing this regime in bulk has proven difficult. Bulk polycrystalline Nd$_{1-x}$Sr$_x$NiO$_2$ obtained by topotactic reduction remains insulating and shows no superconductivity up to pressures of about 50\,GPa~\cite{Li2020absence}. 
Topotactic reduction itself is difficult to control uniformly in single crystals \cite{Puphal2023b}, and chemical doping of perovskite single crystals by divalent alkaline-earth substitution is challenging ~\cite{Puphal2021} and due to a solubility limit ~\cite{Puphal2023} stays below the dome for the reduced IL. 

In this study, we report the single-crystal growth of $n=4$, La$_5$Ni$_4$O$_{13}$ by high-pressure synthesis in a Walker-type multi-anvil module, characterize its structure, perform its topotactic reduction, and demonstrate superconductivity in the resulting square-planar compound La$_5$Ni$_4$O$_{10}$. We systematically explore the growth conditions required to stabilize these high-order RP nickelates under pressure and to topotactically reduce the superconducting quadruple-layer phase. The resulting quadruple-layer nickelate constitutes a bulk, ambient-pressure nickelate superconductor, with a large superconducting volume fraction reaching 60\% and zero-resistance.

\subsection*{ Synthesis of Bulk High-Order RP Nickelates}

The high-order RP nickelate with $n=4$, La$_5$Ni$_4$O$_{13}$, was synthesized in bulk form using the high-pressure high-temperature (HPHT) technique. By subsequently adjusting the La:Ni stoichiometry of the precursor, we also obtained La$_6$Ni$_5$O$_{16}$ ($n=5$) reported in Ref.~\cite{Shi2026} (see Methods and Supplementary Information for details).

Single crystals were extracted from the recovered HPHT pellet and characterized by single-crystal X-ray diffraction (XRD). Reciprocal-space maps of the $(h0l)$ and $(hk0)$ planes (Fig.~\ref{RP}a,b) display sharp, well-separated reflections with no evidence of intergrowth, and structural refinement confirms that the quadruple-layer phase crystallizes in the monoclinic space group $P2/c$, with two crystallographically inequivalent Ni sites corresponding to the outer and inner NiO$_6$ layers of the four-layer block (Fig.~\ref{RP}f; refinement details in Table~\ref{SCtab}). To assess the phase homogeneity of the bulk product, crushed single crystals were further analysed by powder XRD (PXRD). A Rietveld refinement using a single $P2/c$ phase reproduces the observed pattern well (Fig.~\ref{RP}e), confirming that the HPHT synthesis yields a macroscopically homogeneous phase.

We find that pressure is a necessary condition for stabilizing these metastable higher-$n$ phases, but is not itself a fine control parameter for selecting a specific $n$. This observation parallels literature reports on Sr-doped bilayer crystals obtained at both 3 and 20\,GPa, where, notably, the lower-doped composition was synthesized at the higher pressure~\cite{xu2024pressure}, suggesting that pressure primarily provides thermodynamic access to the metastable manifold rather than selecting a unique member within it. The decisive parameter for phase selection is instead the precursor: homogeneous cation mixing, achieved here via sol–gel preparation, is essential for obtaining a single-phase product. By tuning the La:Ni ratio of the sol–gel precursor, we obtained phase-pure batches of either La$_5$Ni$_4$O$_{13}$ or the quintuple-layer La$_6$Ni$_5$O$_{16}$, which crystallizes in the same space group (Fig.~\ref{RP}c,d, Table~\ref{SCtab} and Supplemental Material). The two phases are readily distinguished by PXRD (Fig.~\ref{RP}e): the monoclinic $a$ axis, which runs along the stacking direction, increases from 18.0299(10)~\AA{} to 21.806(2)~\AA{} for $n=4$ to $n=5$. The corresponding stacking period, $a\sin\beta$, grows by \SI{3.8}{\angstrom}, in excellent agreement with the insertion of one additional NiO$_6$ octahedral layer per perovskite block.

\begin{figure*}
    \centering    \includegraphics[width=0.9\linewidth]{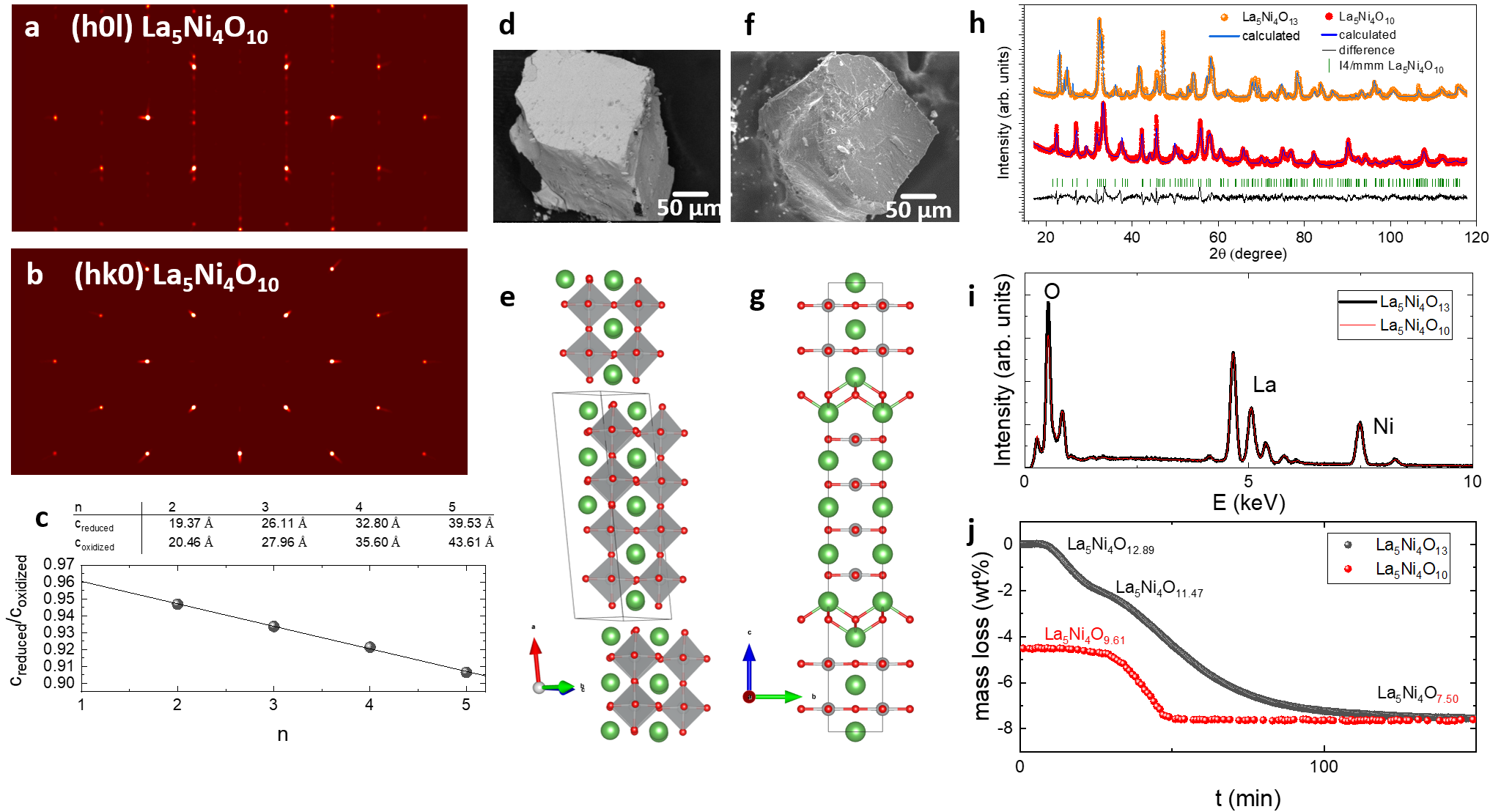}
    \caption{\textbf{Reduction of quadruple-layer nickelates. a,b,} single crystal XRD maps of (h0l), and (hk0) of \LNOr{}. \textbf{c,} relative $c-$axis change by n for $n=2$ \cite{Poltavets2006}, 3 \cite{Poltavets2007}, 4 and 5 revealing a clear linear trend and full reduction for our higher order nickelates. \textbf{d,f,} BSE picture of a \LNO{} and a \LNOr{} single crystal. \textbf{e,g,} Visualization of the crystal structure of \LNO{} and reduced \LNOr{}.  \textbf{h,} PXRD of crushed single crystals with underlying Rietveld refinement of as grown \LNO{} and topotactic reduced \LNOr{} each refined with a single phase. \textbf{i,} EDX spectra on a \LNOr{} in comparison to \LNO{}. \textbf{j} TGA of as grown \LNO{} in comparison to reduced single crystals of \LNOr{}.}
    \label{red}
\end{figure*}

The temperature-dependent resistivity of La$_5$Ni$_4$O$_{13}$ ($n=4$) and La$_6$Ni$_5$O$_{16}$ ($n=5$) is shown in Fig.~\ref{RP}h and Fig.~\ref{RP}j, respectively. Both compounds show a metallic temperature dependence for measurements inplane with the resistance increasing approximately linearly with $T$ for currents majorly running in plane. Around 50\,K a slope change, consistent with defect scattering is seen, which is mirrored by a weak upturn in the susceptibility (Fig.~\ref{RP}i,k) attributable to a Curie-like tail from localized moments associated with oxygen vacancies or trace paramagnetic impurities. For \LNO{} we see reproducibly an additional anomaly around 150 K in transport as well as susceptibility (see Fig.~\ref{RP}h,i), reminiscent of density-wave (DW) transitions found in bilayer and trilayer. Similarly a weak feature might exist around 120 K in \LNOfive{} (see Fig.~\ref{RP}j,k). Note that the weak 220 K anomaly in susceptibility (Fig.~\ref{RP}i,k) possibly originates from Monolayer stacking faults \cite{Reiss2026}. 
We observe the expected trend of increasing metallicity with layer number $n$, consistent with the end-member LaNiO$_3$ ($n\rightarrow\infty$) being a Stoner-enhanced Pauli-paramagnetic metal with a maximum around 220 K \cite{Puphal2023}. We can see a shift of this maximum in susceptibility (Fig.~\ref{RP}i,k) to 300 K in \LNOfive{} and 450 K in  \LNO{}. This evolution is also captured in transport, where equal contact layouts on equal sized samples were performed. We can see a reduction in resistance by a factor of 10 with $n$. 
In the susceptibility: La$_6$Ni$_5$O$_{16}$ displays Pauli-paramagnetic behaviour over a wide temperature range, whereas La$_5$Ni$_4$O$_{13}$ has a more pronounced anomaly in susceptibility around 150 K. This trend falls well in line with our studies on n =1-3, $\infty$ ~\cite{Reiss2026} and the progression suggests that applying pressure to La$_5$Ni$_4$O$_{13}$ is a promising route toward tuning or suppressing its possibly DW-like order, in analogy with the lower-$n$ members.

\subsection*{ Reduction of Quadruple-Layer Nickelates}

\begin{figure*} [h]
    \centering
    \includegraphics[width=0.9\linewidth]{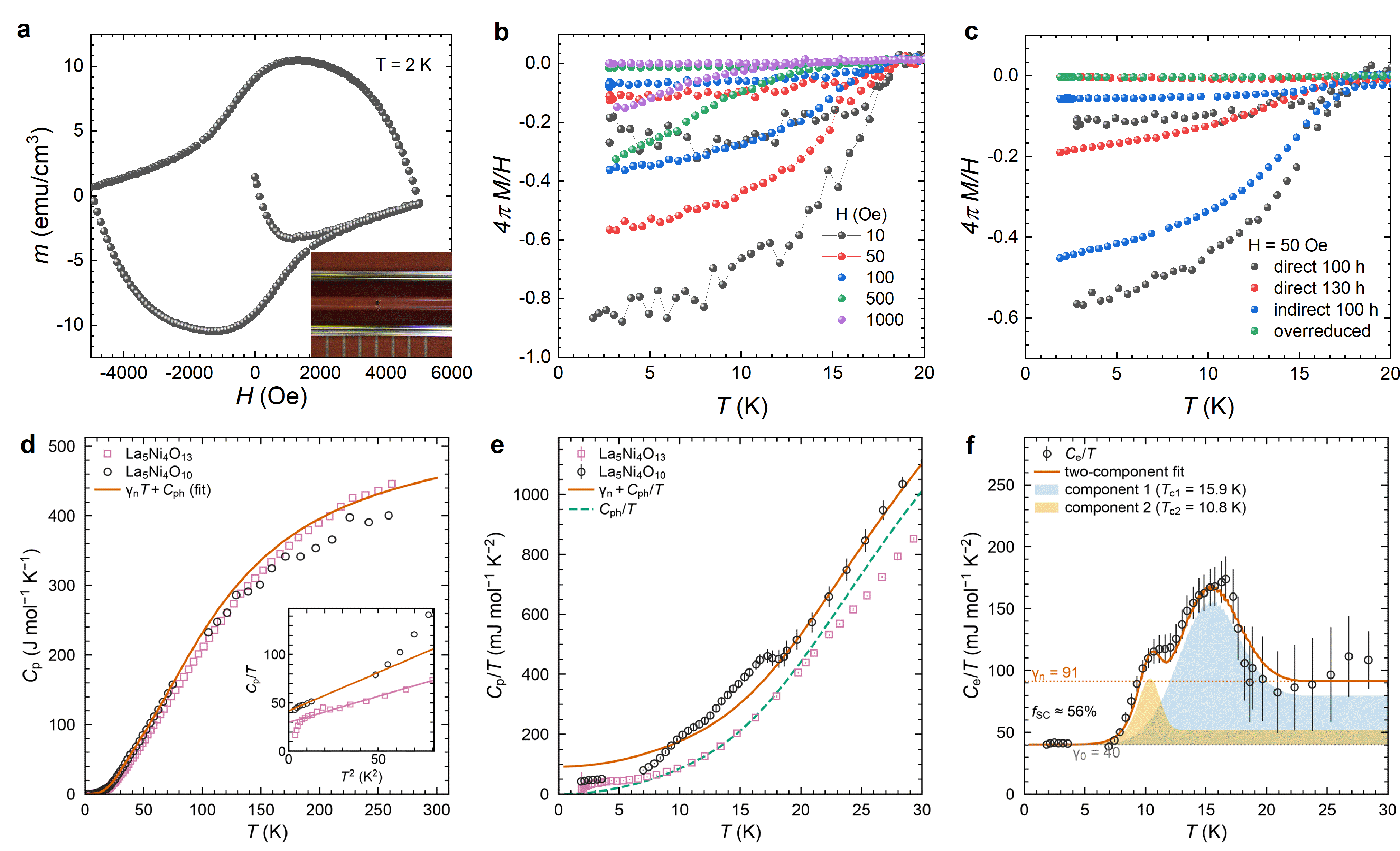}
    \caption{\textbf{Superconducting Volume fraction a,} Field dependent magnetization at 2 K for ZFC and then cycling through -5000 to 5000 Oe. The inset shows the mounted Tetragonal platelets of approximately 200 $\mu$m of \LNOr{} on a quartz SQUID holder, which leads to a small demagnetization factor. \textbf{b,} Temperature dependent magnetization of optimal reduced \LNOr{} measured at varied field in ZFC and FC manner at 10 (black), 50 (red), 100 (blue), 500 (green) and 1000 Oe (purple).  \textbf{c,}  Magnetic screening of \LNOr{} reduced under varying conditions. \textbf{d,e,f} Temperature dependent magnetization measured at an external field of 50 Oe of ampoule 1, 2 and 3 reduced at direct contact for 100h, 130h and indirect contact 100h. Lastly, an overreduced crystal signal is shown (green). \textbf{d},~Specific heat $C_p$ of \LNO{} and \LNOr{} from 2 to 250~K; the dashed line is the
    Dulong--Petit value $3NR$. Inset: $C_p/T$ versus $T^2$ at low temperature with
    a fit to $\gamma_0 + \beta T^2$.
    \textbf{e},~$C_p/T$ below 30~K. The solid line is the normal-state contribution
    $\gamma_{\mathrm{n}} + C_{\mathrm{ph}}/T$, obtained from a Debye-plus-two-Einstein
    phonon model constrained by entropy conservation; the dashed line is the phonon
    part alone.
    \textbf{f},~Electronic specific heat
    $C_{\mathrm{e}}/T = (C_p - C_{\mathrm{ph}})/T$ with a fit to a two-component $\alpha$-model including a residual term $\gamma_0$ (solid line). The shaded areas show the contributions of the two components above $\gamma_0$, each with a Gaussian distribution of $T_{\mathrm{c}}$. The superconducting volume fraction is estimated as $f_{\mathrm{SC}} = 1 - \gamma_0/\gamma_{\mathrm{n}}$.}
    \label{squid}
\end{figure*}

Topotactic reduction was carried out successfully while preserving single crystallinity throughout, as confirmed by the single-crystal XRD maps in Fig.~\ref{red}a,b. Elemental sensitive backscattered electron imaging (BSE) shown in Fig.~\ref{red}d,f reveals homogeneous samples. This outcome contrasts markedly with the reduction of perovskite LaNiO$_3$, where extensive domain formation severely limits the retained crystallinity~\cite{Puphal2021,Puphal2023b}. We note that the incongruent melting behaviour characteristic of higher-order RP phases typically produces finite NiO inclusions in the as-grown crystals~\cite{Puphal2023}, which upon reduction give rise to ferromagnetic Ni metal impurities~\cite{Wu2024}.

Comparing the relative $c$-axis contraction of our reduced La$_5$Ni$_4$O$_{10}$ crystals with literature values for other $n$ (Fig.~\ref{red}c), we find a clear linear reduction trend across the series, where the lower-$n$ members were previously characterized by neutron diffraction~\cite{Poltavets2006,Poltavets2007}. 

Inspecting the structure shown Fig.~\ref{red}e,g we wish to note the migration of rock-salt layer oxygen toward fluorite-type coordination in all reduced. The reduced phase, nominally La$_5$Ni$_4$O$_{10}$, is well fitted in the tetragonal space group $I4/mmm$ by PXRD (Fig.~\ref{red}h; structural details in Table~\ref{SCtab}), consistent with the single-crystal results and confirming a homogeneous reduced phase free of residual higher-oxygen containing impurities. Energy-dispersive X-ray spectroscopy (EDX) confirms an unchanged La:Ni ratio of 5:4 before and after reduction (Fig.~\ref{red}i); the relative change in oxygen content can also be qualitatively tracked from the reduced intensity of the O K$_\alpha$ line at 0.52\,keV. Quantitative EDX analysis of approximately 20 points yields a reduced-phase stoichiometry of La$_{5.0(3)}$Ni$_{3.9(6)}$O$_{9.5(17)}$, consistent within error with full reduction to the $n=4$ square-planar composition.

Thermogravimetric analysis (TGA, Fig.~\ref{red}j) independently corroborates this conclusion. Heating drives both the as-grown and reduced samples toward complete decomposition into $2.5\,\mathrm{La_2O_3} + 4\,\mathrm{Ni}$ (molar mass 1049.29\,g\,mol$^{-1}$). The as-grown sample loses 7.6\,wt\%, corresponding to an oxygen content of $7.5 + (7.6/15.999)/[(100-7.6)/1049.29] = 12.89$, in good agreement with the nominal La$_5$Ni$_4$O$_{13}$ stoichiometry. The reduced sample loses only 3.11\,wt\%, corresponding to $7.5 + (3.11/15.999)/[(100-3.11)/1049.29] = 9.605$. Taken together with the EDX results, these data establish that the reduction proceeds essentially to completion, yielding a fully reduced La$_5$Ni$_4$O$_{10}$ phase that is already approaching the onset of thermal decomposition.

\subsection*{Volume fraction of superconductivity in La$_5$Ni$_4$O$_{10}$}

Superconductivity was observed in reduced La$_5$Ni$_4$O$_{10}$ crystals obtained both by direct and by indirect reduction inside sealed ampoules (see Methods). Remarkably, irrespective of the precise reduction conditions, reduction time, or sample mass, every superconducting crystal displays the same transition temperature, $T_\mathrm{c}\simeq18$\,K, indicating that $T_\mathrm{c}$ is an intrinsic property of the fully reduced phase rather than a parameter sensitive to processing details. Field-dependent magnetization at 2\,K shows a clear hysteresis loop (Fig.~\ref{squid}a), from which we estimate a lower critical field of $H_{c1}\approx1060$\,Oe, close to the value reported for La$_6$Ni$_5$O$_{12}$~\cite{Shi2026}. In the optimally reduced samples, the superconducting volume fraction exceeds 60\%, as estimated from the zero-field-cooled (ZFC) shielding signal (Fig.~\ref{squid}b) at 10\,Oe.

To evaluate the reproducibility of the reduction process, we measured multiple single crystals of comparable size, typically 200$\times$200$\times$100–200\,$\mu$m$^3$ (Supplementary Information), corresponding to a mass of roughly 0.01-0.02\,mg ($\rho=$7.04\,g\,cm$^{-3}$) per crystal. To minimize spurious background contributions, crystals were mounted on quartz holders using a minimal amount of grease (inset of Fig.~\ref{squid}a). We isolated and subtracted the ferromagnetic Ni and diamagnetic background contributions by separately measuring several over-reduced crystals under identical temperature and field conditions (Fig.~\ref{squid}c); because the crystals were measured with the field in the plane of the platelets, the associated demagnetization factor is minimal. Most samples were additionally confirmed by single-crystal XRD to retain the quadruple-layer structure prior to magnetic screening.

For direct-contact reduction over 100\,h (Fig.~\ref{squid}d), all crystals of approximately 0.01\,mg show a consistent ZFC shielding signal at 2\,K of about $-0.65$ (where -1 is expected for perfect diamagnetism); larger crystals show a reduced volume fraction, consistent with the reduction depth being limited by diffusion and therefore sample-volume dependent. Extending the direct-contact reduction time to 130\,h (Fig.~\ref{squid}c) suppresses the shielding response to about $-0.2$ for samples of comparable mass, indicating that prolonged exposure to the reducing agent degrades, rather than improves, the superconducting volume fraction, likely through over-reduction or the onset of decomposition suggested by the TGA data above. Indirect-contact reduction (Fig.~\ref{squid}c) produces larger sample-to-sample variation: a subset of crystals reaches shielding fractions comparable to the best direct-reduced samples, but the majority show substantially weaker signals, reflecting a less controlled and more spatially inhomogeneous nature of vapor-phase CaH$_2$ reduction, possibly showing an even stronger scaling by volume.

Specific heat provides thermodynamic evidence for bulk superconductivity (Fig.~\ref{squid}d--f). $C_p/T$ shows a clear jump at $T_\mathrm{c}$ (Fig.~\ref{squid}d--f), and the electronic specific heat, $C_\mathrm{e}/T = (C_p -C_\mathrm{ph})/T$, rises sharply below about \SI{18.5}{K}, close to the resistive onset (Fig.~\ref{squid}f). The jump, $\Delta C_\mathrm{e}/T_\mathrm{c} \approx \SI{80}{mJ\,mol^{-1}\,K^{-2}}$, is a large fraction of the normal-state Sommerfeld coefficient, $\gamma_\mathrm{n} \approx \SI{91}{mJ\,mol^{-1}\,K^{-2}}$. At low temperatures, $C_p/T = \gamma_0 + \beta T^2$ with $\gamma_0 \approx \SI{41}{mJ\,mol^{-1}\,K^{-2}}$ (Fig.~\ref{squid}d, inset), indicating that part of the sample remains normal. The superconducting part thus contributes $\gamma_\mathrm{n} - \gamma_0 \approx \SI{50}{mJ\,mol^{-1}\,K^{-2}}$, corresponding to a superconducting volume fraction of $f_\mathrm{SC} = 1 - \gamma_0/\gamma_\mathrm{n} \approx 56\%$, in agreement with the magnetic shielding. For this fraction, the weak-coupling BCS ratio $\Delta C_\mathrm{e}/(\gamma T_\mathrm{c}) = 1.43$ predicts a jump of about \SI{70}{mJ\,mol^{-1}\,K^{-2}}, close to the measured value. The size of the jump is therefore consistent with bulk superconductivity in about half of the sample.

The normal-state contribution, $\gamma_\mathrm{n}T + C_\mathrm{ph}$, was determined with a Debye-plus-two-Einstein phonon model constrained by conservation of the electronic entropy (Supplementary Information). As an
independent check of the phonon background, the unreduced La$_5$Ni$_4$O$_{13}$
shows no anomaly in this temperature range, and its $C_p/T$ agrees with the
fitted $C_\mathrm{ph}/T$ within 3\% between 10 and \SI{18}{K}
(Fig.~\ref{squid}e). Varying the fitting ranges changes $f_\mathrm{SC}$
by only a few percent (53--57\%; Supplementary Information).

The transition is broad and shows a main maximum near \SI{16}{K} and a
shoulder near \SI{11}{K} (Fig.~\ref{squid}f). Such a broad transition
likely reflects an inhomogeneous degree of oxygen reduction within the
crystals, consistent with the dependence of the superconducting fraction on the
reduction conditions discussed above. A detailed analysis of the shape of the
anomaly, including a comparison of two-component and two-gap models, is given
in the Supplementary Information.

\subsection*{ Transport of Bulk La$_5$Ni$_4$O$_{10}$ Superconductor}

\begin{figure*}[h]
    \centering
    \includegraphics[width=0.9\linewidth]{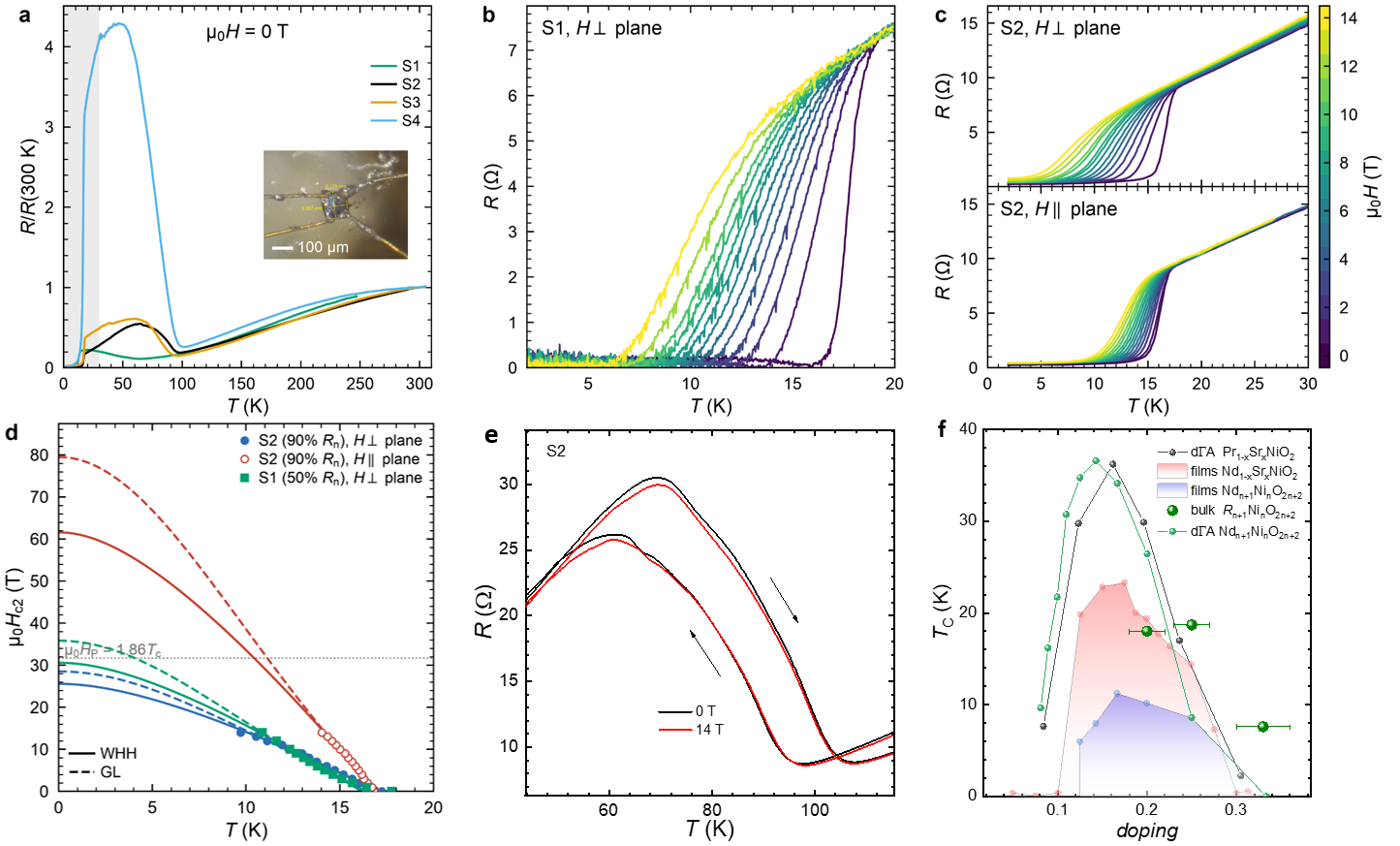}
    \caption{\textbf{Transport and thermodynamic properties of bulk \LNOr{} superconductor}
     \textbf{a}, Temperature dependence of the normalized resistance $R/R(\mathrm{300~K})$ for four samples S1, S2, S3 and S4 at zero field from 300~K to 2~K, with cured epoxy silver paint. Inset: optical micrograph of S2 with four contacts. \textbf{b},~ Temperature dependence of a sample showing zero resistance in various applied fields untreated silver paint. \textbf{c,} Shown is the Low-temperature resistance under magnetic fields of 0--14~T applied
    out of plane (top) and in plane (bottom).
    \textbf{d,}~Upper critical field $\mu_0 H_{\mathrm{c2}}$ determined from the
    90\% $R_{\mathrm{n}}$ criterion for both field orientations (S2) and 50\% $R_{\mathrm{n}}$ criterion for out-of-plane field orientations (S1). Solid and dashed
    lines are fits to the WHH and GL models, respectively; the dotted line marks
    the weak-coupling Pauli limit $\mu_0 H_{\mathrm{P}} = 1.86\,T_{\mathrm{c}}$.
    \textbf{e}, Temperature dependence of the resistance through the 90 K transition for heating and cooling in 0 and 14 T.  \textbf{f}, Doping dependence of the superconducting onset temperature in IL Nd$_{1-x}$Sr$_x$NiO$_2$ \cite{Lee2023}, theoretical d$\Gamma$A studied PR$_{1-x}$Sr$_x$NiO$_2$ \cite{Kitatani2020}, reduced RP Nd$_{n+1}$Ni$_n$O$_{2n+2}$ \cite{Pan2026} and its d$\Gamma$A variant \cite{Hausoel2025}, as well as the recent results of \LNOfiver{} \cite{Shi2026}, our \LNOr{} data and the recent observation in (La,Pr)$_4$Ni$_3$O$_{8}$ \cite{Ahn2026}.
    }
    \label{fig4_bulk}
\end{figure*}

The superconducting transition is clearly resolved in temperature-dependent resistance measurements (Fig.~\ref{fig4_bulk}a-c). All crystals from the same growth and reduction batch exhibit a superconducting transition, although the absolute normal-state resistance varies between individual samples, as expected for crystals of different geometry and residual defect density. Notably, the method for producing electrical contacts on reduced samples also has an effect on the final result. In Fig.~\ref{fig4_bulk}b and c, we compare results using room-temperature drying conductive paint versus cured epoxy: the former shows a (near) zero resistance, whereas the latter shows finite resistance values (Fig.~\ref{fig4_bulk}a-c). On cooling, all samples show a resistive anomaly near \SI{98}{K}, below which the resistance increases, suggestive of a transition into either a partially gapped density-wave state. This transition shows hysteretic behaviour upon heating and cooling, but reveals no field dependence (Fig.~\ref{fig4_bulk}e). The reduced trilayer nickelates show a competing charge-stripe order likely linked with a spin-stripe order \cite{Chen2022a} separating Ni$^{1+}$ and Ni$^{2+}$ raising a plausible scenario for our reduced samples.
This upturn is weak in samples S2 and S3 but much more pronounced in S4, where $R$ increases to roughly four times $R(\SI{300}{K})$; the microscopic origin of this anomaly, and of its pronounced sample-to-sample variability, remains to be clarified, but plausibly reflects differences in local oxygen stoichiometry or strain introduced during reduction. For sample S2, the superconducting onset (90\% of $R_\mathrm{n}$) occurs at $T_\mathrm{c}^\mathrm{onset}=\SI{17.2}{K}$, and the resistance reaches 50\% of $R_\mathrm{n}$ at \SI{16.5}{K}. Below the transition, the resistance drops by about 98\%, leaving a small residual of roughly 2\% of $R(\SI{20}{K})$ down to \SI{2}{K}, which likely stems from surface decomposition of the sample in air.

Magnetic fields suppress superconductivity more strongly when applied perpendicular ($H_\perp$) than parallel ($H_\parallel$) to the NiO$_2$ planes (Fig.~\ref{fig4_bulk}b,c,d), but the anisotropy is small. The ratio $\gamma = H_{c2}^{\parallel}/H_{c2}^{\perp} \approx 2.4$ remains finite near $T_\mathrm{c}$, and $H_{c2}^{\parallel}(T)$ is close to linear: a fit to $H_{c2}^{\parallel}\propto(1-T/T_\mathrm{c})^{n}$ gives $n = 0.85$--0.95, rather than $n = 1/2$ expected for a two-dimensional superconductor. In nickelate thin films, by contrast, the finite thickness leads to $H_{c2}^{\parallel}\propto(1-T/T_\mathrm{c})^{1/2}$ and a diverging anisotropy. The Ginzburg--Landau coherence lengths obtained from the initial slopes, $\xi_{ab}(0) \approx$ \SI{3}{nm} and $\xi_c(0) \approx$ 1.2--\SI{1.8}{nm} depending on the criterion (Supplementary Information), exceed both the spacing of the NiO$_2$ planes within a four-layer block (\SI{0.33}{nm}) and the distance between neighbouring blocks across the La--O spacer (\SI{0.65}{nm}). La$_5$Ni$_4$O$_{10}$ is thus an anisotropic, three-dimensional superconductor in which the four-layer blocks are appreciably coupled along $c$. For in-plane fields, the orbital-limited extrapolation gives $\mu_0H_{c2}^{\parallel}(0)\approx\SI{63}{T}$, well above the weak-coupling Pauli limit of about \SI{32} {T}~\cite{clogston1962upper,chandrasekhar1962note}, which makes \LNOr{} a good candidate to examine the presence or absence of
Pauli-limiting behaviour in high magnetic fields.

Finally, we place the our bulk superconductor \LNOr{}, \LNOfiver{} \cite{Shi2026}, the chemical pressure stabilized $n=3$ \cite{Ahn2026} in a doping dependent phase diagram compared to thin film results on IL \cite{Lee2023}, reduced RP \cite{Pan2026} and their theoretical estimates of the dome \cite{Kitatani2020,Hausoel2025} in Fig. \ref{fig4_bulk}f. Interestingly, combined the recent bulk reports would explain the close lying $T_C$ of $n=4$ and 5 by a relative shift of the dome for bulk samples. Overall, the raised $T_C$ compared to thin films suggests better crystallinity.

\subsection*{ Discussion and Conclusion}

In this work, we overcome the ambient-pressure thermodynamic limit on the perovskite layer number $n$ in RP nickelates through HPHT synthesis. Combined with careful control of precursor homogeneity, this approach yields phase-pure bulk crystals and allows $n$ itself to be used as a synthetic tuning parameter. Topotactic reduction of the quadruple-layer ($n=4$) compound La$_5$Ni$_4$O$_{13}$ yields the square-planar phase La$_5$Ni$_4$O$_{10}$, which we show to be superconducting with $T_\mathrm{c}^\mathrm{onset}\approx18$\,K.

The superconductivity is unambiguously a bulk property of these crystals: transport, magnetization, and specific heat all register the transition at essentially the same temperature, with superconducting volume fractions of about 60\% from zero-field-cooled shielding and 56\% from $1-\gamma_0/\gamma_\mathrm{n}$, two independent estimates that agree remarkably well. The reduction step and possibly air sensitivity appear to be the principal factor currently limiting the superconducting fraction: superconductivity is optimized only within a narrow window of reduction time, depends on sample mass through diffusion-limited kinetics, and is most spatially homogeneous when crystals are in direct physical contact with the reducing agent. The measurement result in three components of the specific-heat anomaly, with $T_\mathrm{c1}\approx16$\,K and $T_\mathrm{c2}\approx11$\,K and a gapped insulating state with finite gamma, which all can be explained by regions of differing local oxygen content (which must then cluster and not randomly distribute possibly due to surface and bulk oxygen contents)  within the crystal rather than by an intrinsic multi-gap pairing state. 

The availability of bulk single crystals also allows the superconducting anisotropy to be measured without the finite-thickness confinement and substrate effects that affect thin films. The anisotropy is small: $\gamma = H_{c2}^{\parallel}/H_{c2}^{\perp} \approx 2.4$ remains finite as $T \rightarrow T_\mathrm{c}$, $H_{c2}^{\parallel}(T)$ is close to linear, and the out-of-plane coherence length exceeds the separation of neighbouring four-layer blocks. Together with the four NiO$_2$ layers per block, this indicates a substantial interlayer coupling across the La--O spacer, making La$_5$Ni$_4$O$_{10}$ an anisotropic but three-dimensional superconductor. 

More broadly, these results establish the reduced RP series $R_{n+1}$Ni$_n$O$_{2n+2}$ as a viable bulk platform for nickelate superconductivity at ambient pressure. In this family, the layer number $n$ simultaneously controls the electronic dimensionality and the nominal Ni valence, $1+1/n$, entirely without chemical doping, a self-doping mechanism with no direct analogue in the cuprates. Now that HPHT synthesis has lifted the long-standing restriction on $n$ in bulk RP nickelates, the series can be explored systematically in both directions: toward larger $n$, to map out the evolution of $T_\mathrm{c}$ and dimensionality as the system approaches the IL limit, and through rare-earth substitution at fixed $n$, to introduce chemical pressure or additional doping as a further means of enhancing $T_\mathrm{c}$. The availability of bulk single crystals, rather than films or diamond-anvil-cell samples, also opens the door to probes that are difficult or impossible to apply otherwise: inelastic neutron scattering to map the spin and lattice dynamics, muon spin rotation to probe the superfluid density and pairing symmetry, and angle-resolved photoemission on cleaved surfaces. Such measurements should help to clarify the microscopic pairing mechanism in the nickelates and to position this emerging family of superconductors within the broader landscape of unconventional superconductivity alongside the cuprates and iron-based materials. The coexistence of the superconductivity with an anisotropic charge order or MIT around 90 K in all our samples shows that the compound resides in the lower doping regime of the dome and reveals clear analogy to cuprates. The closely related $T_C$ from \LNOr{} and \LNOfiver{} could indicate a similar oxidation state either from partial reduction in either case, or from charge order that could exist in both samples, while a significantly lower $T_C$ is observed for stacking faulted (La,Pr)$_4$Ni$_3$O$_8$ \cite{Ahn2026}.

\begin{table*}[htbp]
\centering
\caption{Lattice parameters and refined atomic coordinates for
\LNO{}, \LNOr{}, and \LNOfive{}  from single crystal XRD results with the corresponding maps shown in Fig.~\ref{RP} and ~\ref{red}.}
\label{tab:structural-parameters}
\resizebox{\textwidth}{!}{%
\begin{tabular}{l|cccc|cccc|cccc}
\toprule
& \multicolumn{4}{c}{\LNO{} (\#13 $P2/c$)}
& \multicolumn{4}{c}{\LNOr{} (\#139 $I4/mmm$)}
& \multicolumn{4}{c}{\LNOfive{}(\#14 $P2_1/c$)} \\
\cmidrule(lr){2-5}
\cmidrule(lr){6-9}
\cmidrule(lr){10-13}
& $a$ (\AA) & $b$ (\AA) & $c$ (\AA) & $\beta$ ($^\circ$)
& $a$ (\AA) & $b$ (\AA) & $c$ (\AA) & $\beta$ ($^\circ$)
& $a$ (\AA) & $b$ (\AA) & $c$ (\AA) & $\beta$ ($^\circ$) \\
\midrule
& 18.0299(10) & 5.4494(3) & 5.4234(4) & 98.635(6)
& 3.9608(2) & 3.9608(2) & 32.752(3) & --
& 21.806(2) & 5.4546(6) & 5.4020(7) & 96.773(12) \\
\midrule
Site
& $x$ & $y$ & $z$ & $U_{\mathrm{iso}}$
& $x$ & $y$ & $z$ & $U_{\mathrm{iso}}$
& $x$ & $y$ & $z$ & $U_{\mathrm{iso}}$ \\
\midrule
La01
& 0.5 & 1 & 0.5 & 0.0025(4)
& 1 & 1 & 0.39586(6) & 0.0047(7)
& 0.73596(6) & 0.2492(3) & 0.6098(2) & 0.0056(4) \\

La02
& 0.28651(5) & 0.5007(2) & 0.89237(17) & 0.0038(4)
& 1 & 0 & 0.5 & 0.0042(8)
& 0.91192(5) & 0.2505(3) & 0.7042(2) & 0.0041(4) \\

La03
& 0.08011(6) & 0.0006(2) & 0.2893(2) & 0.0075(4)
& 1 & 0 & 0.28942(7) & 0.0086(7)
& 0.43405(6) & 0.7488(3) & 0.4702(2) & 0.0071(4) \\

Ni04
& 0.39225(11) & 0.5007(4) & 0.4456(3) & 0.0002(6)
& 0.5 & 0.5 & 0.44949(14) & 0.0039(11)
& 0.82258(12) & 0.7460(5) & 0.6579(5) & 0.0030(7) \\

Ni05
& 0.17416(15) & 0.4972(5) & 0.3364(4) & 0.0036(6)
& 0.5 & 0.5 & 0.34954(17) & 0.0076(11)
& 0.64348(15) & 0.7518(6) & 0.5674(5) & 0.0047(6) \\

O06
& 0.5 & 0.5 & 0.5 & 0.008(4)
& 1 & 0.5 & 0.4494(5) & 0.007(4)
& 1 & 0.7620(8) & 0.75 & 0.0014(8) \\

O07
& 0.2836(9) & 0.497(3) & 0.392(3) & 0.016(3)
& 1 & 0.5 & 0.3486(6) & 0.008(4)
& 0.6452(7) & 0.495(3) & 0.823(3) & 0.001 \\

O08
& 0.1754(10) & 0.245(3) & 0.066(3) & 0.021(4)
& 1 & -0.5 & 0.25 & 0.005(5)
& 0.7323(9) & 0.758(4) & 0.605(4) & 0.025(5) \\

O09
& 0.1726(8) & 0.742(2) & 0.100(2) & 0.008(3)
& -- & -- & -- & --
& 0.9088(9) & 0.711(4) & 0.701(3) & 0.019(5) \\

O10
& 0.3914(12) & 0.735(3) & 0.190(4) & 0.033(5)
& -- & -- & -- & --
& 1 & 0.5 & 0.5 & 0.005(4) \\

O11
& 0.3932(10) & 0.236(3) & 0.201(3) & 0.024(4)
& -- & -- & -- & --
& 0.5429(8) & 0.718(3) & 0.513(3) & 0.013(4) \\

O12
& 0.0570(16) & 0.521(3) & 0.278(4) & 0.039(6)
& -- & -- & -- & --
& 0.6345(12) & 1.010(4) & 0.312(4) & 0.025(5) \\
O13 & -- & -- & -- & --
& -- & -- & -- & --
& 1 & 1 & 0.5 & 0.084(18) \\
O14 & -- & -- & -- & --
& -- & -- & -- & --
& 0.8232(8) & 0.992(3) & 0.913(3) & 0.008(4) \\
O15& -- & -- & -- & --
& -- & -- & -- & --
& 0.816(2) & 0.517(8) & 0.393(8) & 0.096(15) \\
\bottomrule
\label{SCtab}
\end{tabular}%
}
\end{table*}

\section*{ Methods}

\subsection*{ Precursor synthesis}
Polycrystalline precursors were prepared by a sol–gel route. La$_2$O$_3$ and Ni(NO$_3$)$_2\cdot$6H$_2$O, in the La:Ni molar ratio 6:5, were dissolved in a 1:1 (by volume) mixture of concentrated HNO$_3$ and deionized water. Citric acid was added at twice the mass of the targeted precursor (for example, 5\,g of citric acid for 2.5\,g of precursor). The solution was heated at 120\,$^\circ$C until the water had evaporated and a gel had formed. The temperature was then raised slowly to 230\,$^\circ$C (a rate chosen to moderate the exothermic auto-combustion of the gel) and held overnight. The resulting ash-like powder was ground for 1\,h and heated in flowing oxygen to 800\,$^\circ$C at 50\,$^\circ$C\,h$^{-1}$, with a 12\,h dwell, to fully remove residual organic components. The product was ground a second time for 1\,h prior to use. For the \LNOfive the same procedure was applied with a molar ratio of La:Ni of 7:6, contrasting results of Ref. \cite{Shi2026}.

\subsection*{High-pressure crystal growth}
Crystals of La$_5$Ni$_4$O$_{13}$ were grown in a Walker-type multi-anvil module (Voggenreiter) at 7\,GPa, using an 18/11 octahedral pressure assembly (18\,mm octahedron edge length, 11\,mm anvil truncation edge length) and a LaCrO$_3$ heater, which allowed temperatures sufficient to melt the La–Ni–O system to be reached. The precursor, prepared with a slightly Ni-rich La:Ni ratio of 6:5 relative to the target phase to promote crystal growth from the melt, was sealed in a Pt capsule. After the target pressure was reached, the sample was heated to 1600\,$^\circ$C at approximately 100\,$^\circ$C\,min$^{-1}$, then cooled uniformly to 1480\,$^\circ$C over 8\,h, at which point it was quenched to room temperature; the pressure was subsequently released slowly to ambient conditions. Single crystals were extracted from the recovered pellet under an optical microscope. We additionally tested the sensitivity of the growth to pressure and precursor composition: lowering the synthesis pressure to 5\,GPa, and separately adjusting the precursor stoichiometry to a La:Ni ratio of 5:4 by the addition of La$_2$O$_3$. The lower pressure produced only minor changes in crystallinity while the product phase remained La$_5$Ni$_4$O$_{13}$, in case of added La$_2$O$_3$ we find trilayer impurities as discussed in the Supplementary Information. In contrast, preparing a new precursor with a La:Ni ratio of 7:6 yielded phase-pure five-layer La$_6$Ni$_5$O$_{16}$ crystals, demonstrating that precursor stoichiometry, rather than pressure, is the dominant parameter controlling which RP member is obtained.

\subsection*{ Topotactic reduction}
Single crystals selected from the high-pressure product were reduced using CaH$_2$ in evacuated, sealed quartz ampoules~\cite{Puphal2021}. Each ampoule contained \SI{200}{mg} of CaH$_2$, and three reduction conditions were compared. In ampoules 1 and 2, crystals were scattered directly onto the CaH$_2$ powder, placing them in direct physical contact with the reducing agent. In ampoule 3, crystals were spatially separated from the CaH$_2$, allowing reduction to proceed only via the gas phase (indirect reduction). All ampoules were evacuated to $10^{-6}$\,mbar, sealed, and annealed at \SI{300}{\celsius} for \SI{100}{h} (ampoules 1 and 3) or \SI{130}{h} (ampoule 2).

\subsection*{Diffraction}
Powder XRD data were collected on a Rigaku Miniflex diffractometer in Bragg–Brentano geometry using Cu K$_\alpha$ radiation with a Ni filter. Rietveld refinements were performed using TOPAS V7 software~\cite{coelho2018topas}. Single-crystal XRD was performed on a Rigaku XtaLAB mini II diffractometer using Mo K$_\alpha$ radiation. Data were processed with CrysAlis(Pro), and final structure refinements were carried out using Olex2 with SHELX.

\subsection*{ Transport measurements}
Transport measurements were carried out in a Quantum Design Physical Property Measurement System (PPMS, \SI{14}{\tesla}) using a four-probe van der Pauw configuration. Electrical contacts were made with silver epoxy (EPO-TEK H20E), cured at \SI{350}{\celsius} under argon for 10 minutes. Resistance was measured on cooling under fixed magnetic fields of 0–14\,T, applied either perpendicular or parallel to the NiO$_2$ planes.

\subsection*{ Magnetization measurements}
Magnetic susceptibility measurements were performed on single crystals using a Quantum Design MPMS3 SQUID magnetometer. Specimens were mounted on a quartz holder using a minimal amount of grease (inset of Fig.~\ref{squid}c). Temperature-dependent magnetization, $M(T)$, was measured in direct-current (DC) mode between 1.8 and 300\,K, and between 300 and 800\,K using the oven option. Field-dependent magnetization, $M(H)$, was measured in vibrating sample magnetometer (VSM) mode.

\subsection*{ Specific heat measurements}
Specific heat was measured using the heat-capacity option of a PPMS, which employs the thermal-relaxation technique. For the superconductivity measurements, several single crystals were placed together on the sapphire sample platform using Apiezon N grease. The addenda heat capacity (platform and grease) was measured separately over the same temperature range and subtracted from the raw data. Each relaxation curve was analysed using the two-$\tau$ model~\cite{hwang1997measurement}. Data were collected in zero applied field between 1.8 and 300\,K, with every temperature point measured three times and the results averaged.

\bibliography{bib}
\clearpage 
\onecolumn

\renewcommand\thefigure{SM\arabic{figure}}
\renewcommand\thetable{SM\Roman{table}}
\section*{Supplemental Material}

\subsection*{Synthesis control of \LNO, and \LNOfive}

\begin{figure}
    \centering
    \includegraphics[width=1\linewidth]{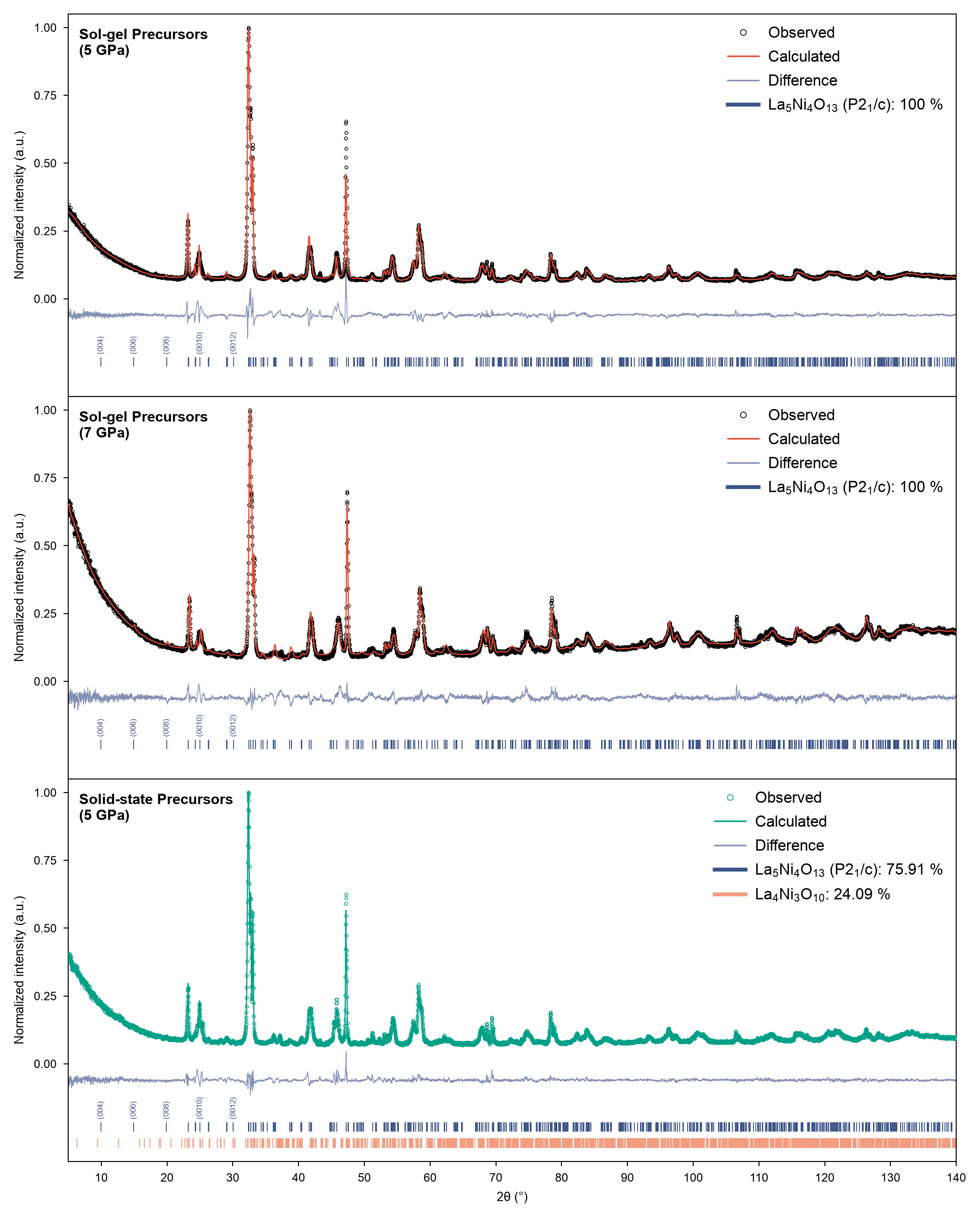}
    \caption{\textbf{Synthesis-condition dependence of phase purity of La$_5$Ni$_4$O$_{13}$.}
    Sol-gel precursors yield single-phase La$_5$Ni$_4$O$_{13}$ ($P2_1/c$) at both pressures, whereas the solid-state precursor yields 75.91\% La$_5$Ni$_4$O$_{13}$ and 24.09\% La$_4$Ni$_3$O$_{10}$ at 5 GPa.}
    \label{SI_figure:homogeniety}
\end{figure}

We synthesized three batches of \LNO{}: two from the sol-gel precursor (see Methods) at 5~GPa and 7~GPa, and one solid-state batch at 5~GPa. For the solid-state batch, we mixed La$_2$O$_3$ into the sol-gel precursor to reach a La:Ni ratio of 5:4; all other conditions were kept the same. Rietveld refinement of the PXRD patterns shows that raising the pressure from 5 to 7~GPa does not introduce any additional phase. The solid-state batch, however, contains \LNOthree{} as a secondary phase. This agrees with our single-crystal selection: in the solid-state batch about one quarter of the crystals were \LNOthree{} and the rest were \LNO{}, whereas all crystals from the sol-gel batches were \LNO{}. We attribute the difference to the poorer homogeneity of the mixed precursor compared with the sol-gel one.

\subsection*{Transport data on \LNO{} and \LNOfive{}}
\begin{figure}
    \centering
    \includegraphics[width=1\linewidth]{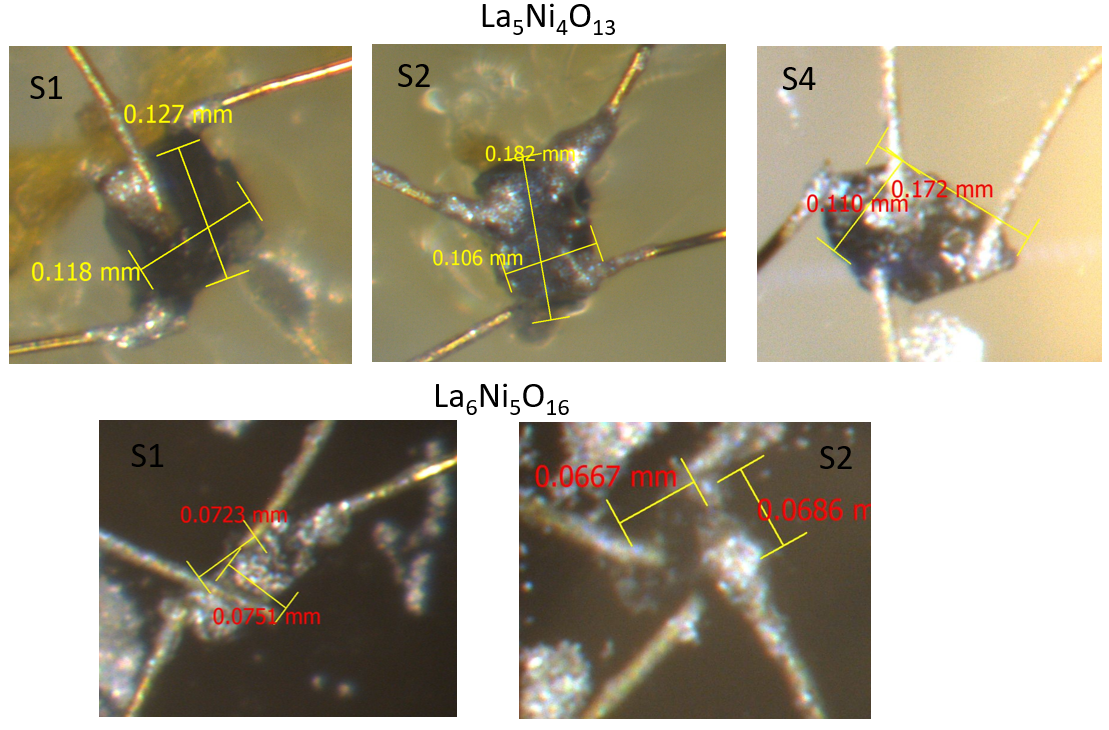}
    \caption{\textbf{Images of the contact layout on \LNO{} and \LNOfive{}.}}
    \label{SI_R}
\end{figure}

In Fig. \ref{SI_R} we show the contacts on the single crystal samples used in the main text to determine the resistivity.

\subsection*{Reduction conditions}
\begin{figure}
    \centering
    \includegraphics[width=1\linewidth]{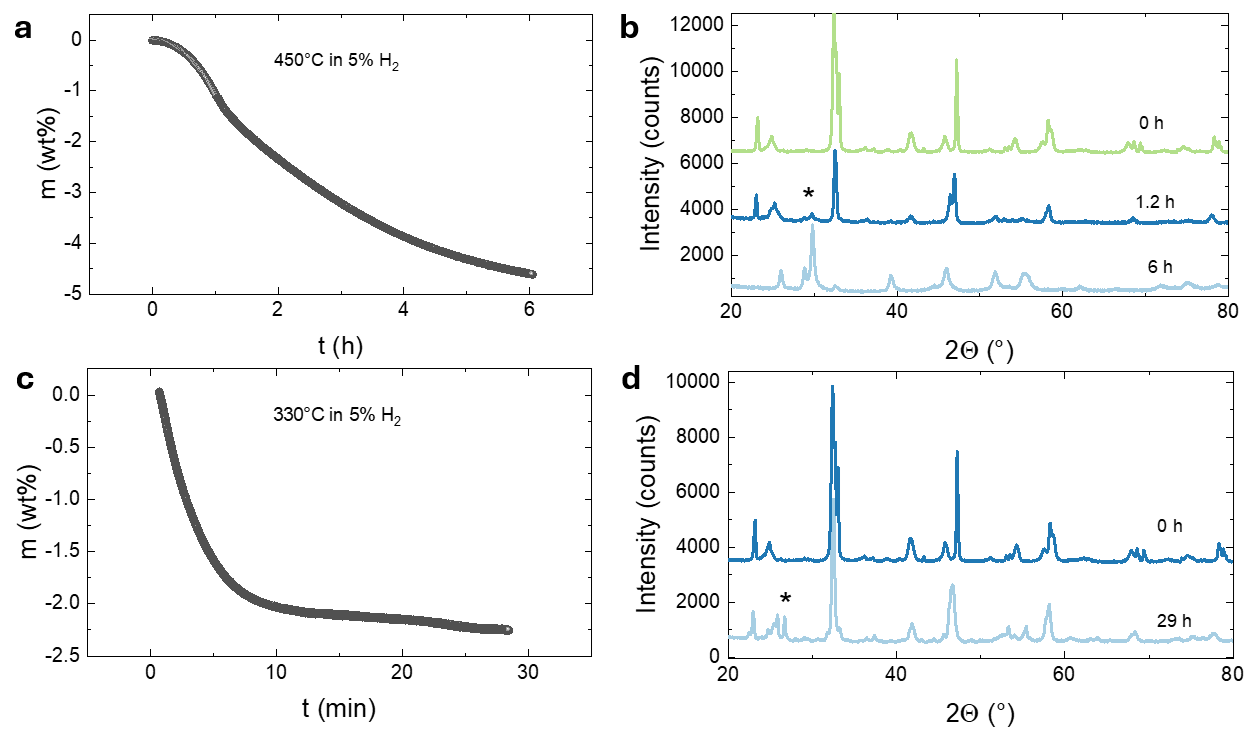}
    \caption{\textbf{Reduction of La$_5$Ni$_4$O$_{10}$ in forming gas.}  \textbf{a,b}, Time dependence of the mass loss of the reduction at 450$^\circ$C and 330$^\circ$C \textbf{c,d} and the corresponding PXRDs with the time given at the curve. Stars denote decomposition phases.}
    \label{SI_figure:reduction}
\end{figure}

Prior to reduction in ampules we tested the reduction of \LNO{} in forming gas displayed in Fig. \ref{SI_figure:reduction}. Here, we found that at 450$^\circ$C we can reach an intermediate La$_5$Ni$_4$O$_{11.5}$ state that is already evident as an intermediate plateau in the full reduction TGA of the main text. Even with a very mild treatment in 330$^\circ$C in forming gas some minor impurity phase developed, which we highligh by a star. We thus turned to CaH$_2$ reduction as discussed in the main text.

\subsection*{Reproducibility}
\begin{figure}
    \centering
    \includegraphics[width=1\linewidth]{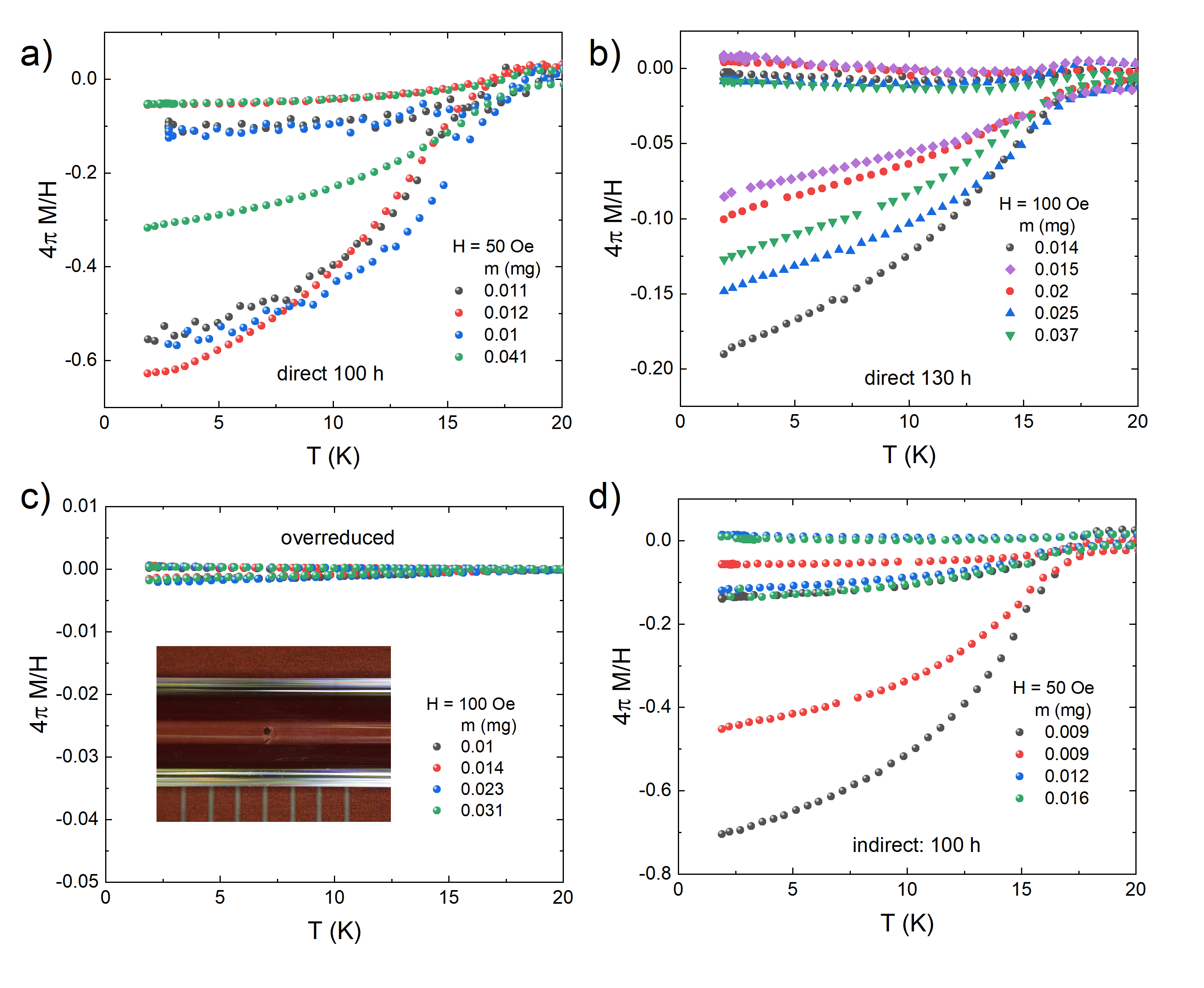}
    \caption{\textbf{Reproducibility of the shielding fractions in each batch}  Temperature dependent magnetization of optimal reduced \LNOr{} measured at constant field of 50/ 100 Oe in ZFC and FC manner for selected crystals of \textbf{a,} ampule 1, \textbf{b,} ampule 2, \textbf{d,} ampule 3 as well as overreduced crystals \textbf{c}.}
    \label{SI_squid}
\end{figure}

We investigated the squid response on a large quantity of samples. In figure \ref{SI_squid} we show > 4 samples for each type of reduction. For the direct contact one in 100 h we see a reproducible shielding fraction of 60\% for equal sized crystals of 0.01 mg, while larger crystals reveal a halfed signal (Fig. \ref{SI_squid}a). Longer reduction time (ampule 2) reduced the superconducting volume fractions to 20\%. In ampule 3 we find some crystals with the largest volume fraction as the method is softer, but the average shows a majorly reduced fraction (see Fig. \ref{SI_squid} d).

\subsection*{Upper-critical field fitting}

\begin{figure}
    \centering
    \includegraphics[width=1\linewidth]{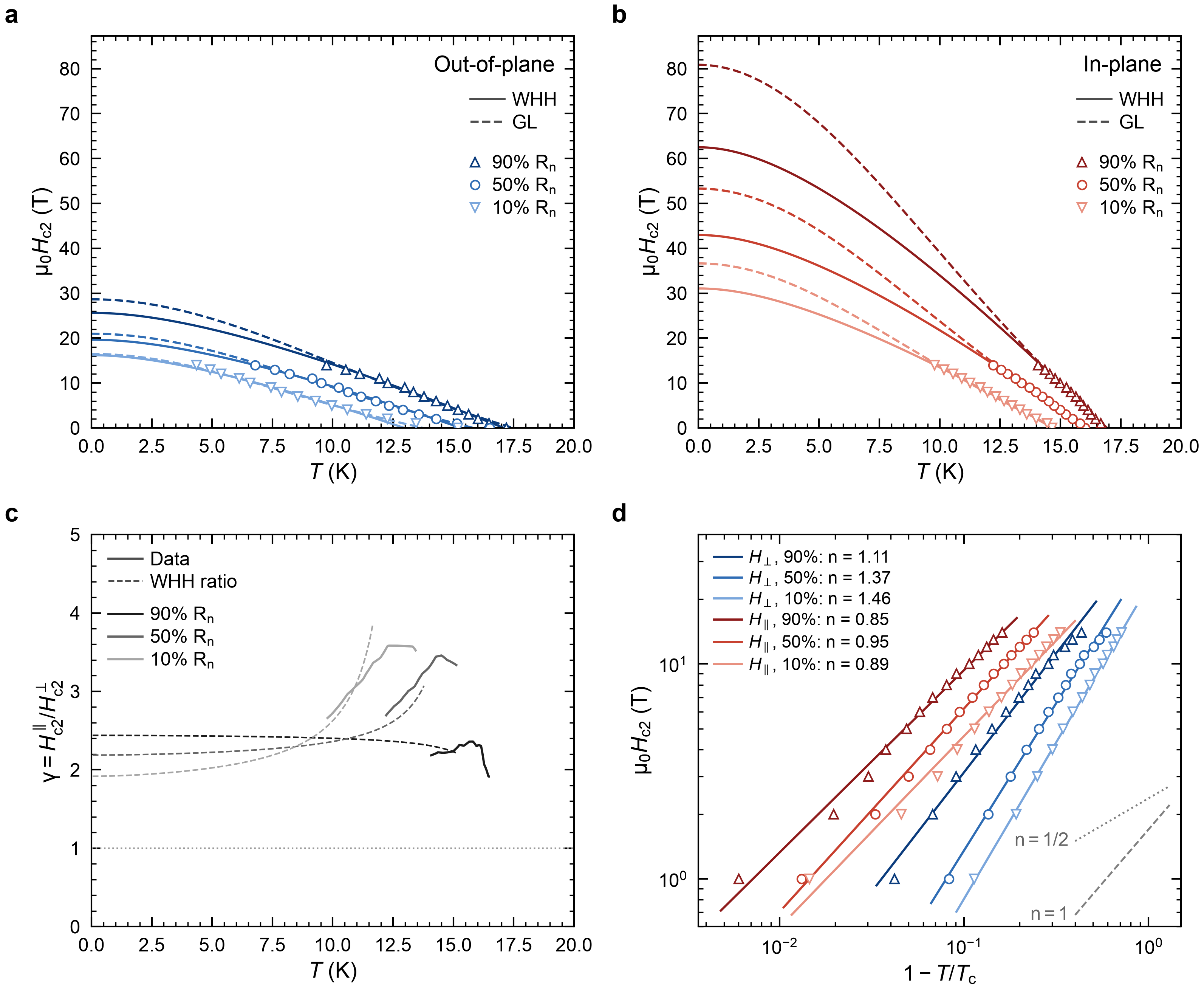}
    \caption{\textbf{Anisotropic upper critical field of La$_5$Ni$_4$O$_{10}$.}  \textbf{a},\textbf{b}, Temperature dependence of the upper critical field   $\mu_0H_{c2}$ for magnetic fields applied out-of-plane (\textbf{a}, $H_\perp$)   and in-plane (\textbf{b}, $H_\parallel$). Solid and dashed lines are fits to the WHH and GL models, respectively. \textbf{c}, Anisotropy ratio $\gamma = H_{c2}^{\parallel}/H_{c2}^{\perp}$. Solid lines are obtained from the experimental $T_c(H)$ data interpolated onto a common temperature grid, restricted to fields $\geq 1$~T. Dashed lines are the ratio of the corresponding WHH fits, shown up to $0.9\,T_c$. \textbf{d}, $\mu_0H_{\mathrm{c2}}$ versus $1 - T/T_{\mathrm{c}}$ on logarithmic axes, with $T_{\mathrm{c}}$ fixed to the zero-field value of each run. Solid lines are power-law fits $\mu_0H_{\mathrm{c2}} \propto (1 - T/T_{\mathrm{c}})^{n}$; grey lines indicate the slopes $n = 1/2$ (two-dimensional) and $n = 1$ (three-dimensional).}
    \label{SI_figure:anistropy_field}
\end{figure}

Because the resistance remains finite below the transition, each curve was normalized as

\begin{equation}
  r(T) = \frac{R(T) - R_0}{R_n(T) - R_0},
\end{equation}

where $R_n(T)$ is a linear fit to the normal-state resistance between 20 and 28~K, extrapolated into the transition region, and $R_0$ is the median resistance within 2~K of base temperature. The transition temperature $T_c(H)$ was defined as the highest temperature at which $r$ crosses 0.9, 0.5 or 0.1 ($90\%$, $50\%$ and $10\%\,R_n$ criteria). 

The temperature dependence of $H_{c2}$ was fitted with two models. The first is the Werthamer--Helfand--Hohenberg (WHH) theory \cite{werthamer1966temperature} in the dirty limit, including only orbital pair breaking ($\alpha = \lambda_{\mathrm{so}} = 0$):
\begin{equation}
  \ln\frac{1}{t} = \psi\!\left(\frac{1}{2} + \frac{\bar h}{2t}\right)
                 - \psi\!\left(\frac{1}{2}\right), \qquad
  \bar h = \frac{4}{\pi^2}\,
           \frac{H_{c2}}{-T_c\,(\mathrm{d}H_{c2}/\mathrm{d}T)_{T_c}},
\end{equation}

where $t = T/T_c$ and $\psi$ is the digamma function. The free parameters are $T_c$ and the initial slope $S = -(\mathrm{d}\mu_0H_{c2}/\mathrm{d}T)_{T_c}$, and the zero-temperature orbital limit is $\mu_0H_{c2}^{\mathrm{orb}}(0) = 0.693\,T_c S$. The second is the empirical Ginzburg--Landau (GL) form $\mu_0H_{c2}(T) = \mu_0H_{c2}(0)\,(1 - t^2)/(1 + t^2)$. 

To test the dimensionality independently of these models, we fitted
\begin{equation}
  \mu_0H_{\mathrm{c2}}(T) = H_0\left(1 - \frac{T}{T_{\mathrm{c}}}\right)^{n},
\end{equation}
with $T_{\mathrm{c}}$ fixed to the measured zero-field transition temperature
and only fields $\geq \SI{1}{T}$ included. For a thin film or a weakly coupled
layered superconductor whose thickness (or interlayer spacing) is smaller than
the out-of-plane coherence length, $H_{\mathrm{c2}}^{\parallel}$ follows the
two-dimensional form with $n = 1/2$; for an anisotropic three-dimensional
superconductor, both orientations follow the linearized GL form with $n = 1$.

\paragraph{Results}
For all three criteria, superconductivity is suppressed more strongly for $H_\perp$ than for $H_\parallel$ (Supplementary Fig.~\ref{SI_figure:anistropy_field}a,b). For the $90\%$ criterion, the initial slopes are $S_\perp = \SI{2.2}{T\per K}$ and $S_\parallel = \SI{5.3}{T\per K}$, and the orbital-limited WHH (GL) fits extrapolate to $\mu_0H_{\mathrm{c2}}^{\perp}(0) = \SI{25.6}{T}$ (\SI{28.6}{T}) and $\mu_0H_{\mathrm{c2}}^{\parallel}(0) = \SI{62.5}{T}$ (\SI{80.9}{T}). The lower criteria give systematically smaller values, down to $\mu_0H_{\mathrm{c2}}^{\perp}(0) = \SI{16.2}{T}$ and $\mu_0H_{\mathrm{c2}}^{\parallel}(0) = \SI{31.1}{T}$ (WHH, $10\%$ criterion).

\paragraph{Dimensionality}The power-law analysis (Supplementary Fig.~\ref{SI_figure:anistropy_field}d)
gives $n = 0.85$--$0.95$ for $H_\parallel$ across the three criteria, close to
the three-dimensional value $n = 1$ and far from the two-dimensional value
$n = 1/2$. Equivalently, the shift of $T_{\mathrm{c}}$ for $H_\parallel$ is
nearly proportional to the field over 0--\SI{14}{T}
($\approx 0.1$--\SI{0.3}{K\per T}), whereas the two-dimensional form would
require $\Delta T_{\mathrm{c}} \propto H^2$. For $H_\perp$, $n = 1.1$--$1.5$,
reflecting a rapid suppression of $T_{\mathrm{c}}$ at low fields
(0.7--\SI{1.7}{K} between 0 and \SI{1}{T}, depending on the criterion). Such
upward curvature of $H_{\mathrm{c2}}^{\perp}(T)$ near $T_{\mathrm{c}}$ is
commonly associated with vortex-liquid or fluctuation broadening in layered
superconductors and becomes more pronounced for lower resistive criteria.

\paragraph{Anisotropy}
The anisotropy ratio obtained from the interpolated data (Supplementary Fig.~\ref{SI_figure:anistropy_field}c) is $\gamma \approx 1.9$--$2.4$ for the $90\%$ criterion and $\approx 2.7$--$3.6$ for the $50\%$ and $10\%$ criteria. From the initial slopes, the GL coherence lengths are $\xi_{ab}(0) = [\Phi_0/(2\pi T_{\mathrm{c}} S_\perp)]^{1/2} \approx \SI{3.0}{nm}$ and $\xi_c(0) = \xi_{ab}(0)/\gamma \approx \SI{1.2}{nm}$ ($90\%$ criterion), where $\Phi_0$ is the flux quantum. In superconducting nickelate thin films, the effective superconducting thickness extracted from two-dimensional GL fits is comparable to the film thickness (4--\SI{8}{nm}), and $H_{\mathrm{c2}}^{\parallel}$ follows $(1 - T/T_{\mathrm{c}})^{1/2}$ near $T_{\mathrm{c}}$, so that $\gamma$ diverges at $T_{\mathrm{c}}$~\cite{liu2025superconductivity,hsu2026fermi,wang2026pauli}; the anisotropies reported for these films ($\gamma \approx 1.3$--$2.5$) are zero-temperature extrapolations. In the bulk crystal studied here, no such geometric confinement exists, and the finite $\gamma$ together with $n \approx 0.85$--$0.95$ for $H_\parallel$ indicates anisotropic three-dimensional superconductivity with coherent interlayer coupling. An angle-dependent measurement of $H_{\mathrm{c2}}(\theta)$ near $T_{\mathrm{c}}$, which would distinguish the smooth three-dimensional GL form from the cusp of the two-dimensional Tinkham model, would provide a further test.


\paragraph{Uncertainties}
The values of $H_{\mathrm{c2}}(0)$ are extrapolations from data limited to \SI{14}{T}; the spread between the WHH and GL results and between the resistive criteria represents the dominant systematic uncertainty, which is much larger than the statistical fit errors ($\lesssim \SI{2}{T}$). The zero-field $T_{\mathrm{c}}$ differs by about \SI{0.4}{K} between the two cooldowns ($90\%$ criterion: \SI{17.19}{K} for the $H_\perp$ run and \SI{16.78}{K} for the $H_\parallel$ run), which mainly affects $\gamma(T)$ close to $T_{\mathrm{c}}$; the slope ratio and the exponents $n$ are not affected, because $T_{\mathrm{c}}$ is fixed to the zero-field value of each run. Finally, any misalignment between the field and the NiO$_2$ planes reduces the apparent anisotropy, so the measured $\gamma$ should be regarded as a lower
bound.

\subsection*{Specific Heat}

\begin{figure}[p]
  \centering
  \includegraphics[width=\linewidth,height=0.8\textheight,keepaspectratio]{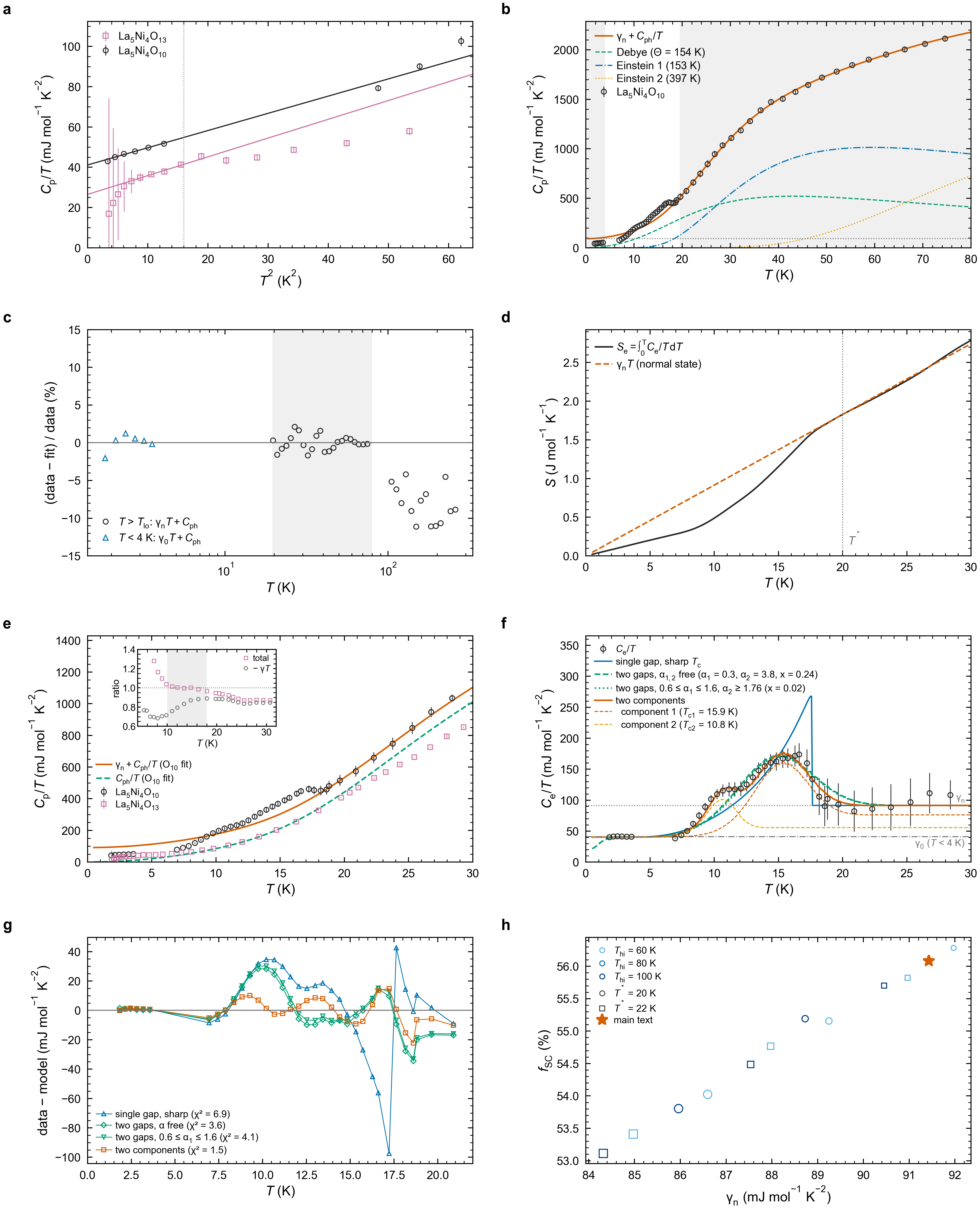}
  \caption{\textbf{Specific heat of La$_5$Ni$_4$O$_{10}$.}
  \textbf{a},~$C_p/T$ versus $T^2$ for La$_5$Ni$_4$O$_{10}$ (circles) and
  La$_5$Ni$_4$O$_{13}$ (squares), with fits to $\gamma + \beta T^2$ below \SI{4}{K}.
  \textbf{b},~Normal-state fit $\gamma_{\mathrm{n}} + C_{\mathrm{ph}}/T$ with its
  Debye and Einstein terms; shading marks the fitting ranges.
  \textbf{c},~Relative residuals of the fit in \textbf{b}.
  \textbf{d},~Electronic entropy compared with $\gamma_{\mathrm{n}}T$.
  \textbf{e},~$C_p/T$ of both compounds with the phonon background of
  La$_5$Ni$_4$O$_{10}$. Inset: ratio of La$_5$Ni$_4$O$_{13}$ data to $C_{\mathrm{ph}}$.
  \textbf{f},~$C_{\mathrm{e}}/T$ with fits to the single-gap, two-gap and
  two-component $\alpha$ models.
  \textbf{g},~Fit residuals.
  \textbf{h},~$\gamma_{\mathrm{n}}$ and $f_{\mathrm{SC}}$ for different fitting
  ranges; the star marks the values used in the main text.}
  \label{fig:SI_HC}
\end{figure}

\paragraph{Low-temperature specific heat.}
Below \SI{4}{K}, $C_p/T = \gamma_0 + \beta T^2$ gives
$\gamma_0 = 41.1 \pm \SI{0.5}{mJ\,mol^{-1}\,K^{-2}}$ and
$\beta = 0.86 \pm \SI{0.06}{mJ\,mol^{-1}\,K^{-4}}$ for La$_5$Ni$_4$O$_{10}$,
corresponding to
$\Theta_{\mathrm{D}} = (12\pi^4 NR/5\beta)^{1/3} = 351 \pm \SI{9}{K}$
(Supplementary Fig.~\ref{fig:SI_HC}a). The finite $\gamma_0$ in the
superconducting state indicates a residual density of ungapped
quasiparticles. For La$_5$Ni$_4$O$_{13}$, the same fit gives
$\gamma = 26 \pm \SI{1}{mJ\,mol^{-1}\,K^{-2}}$ and
$\beta = 0.93 \pm \SI{0.07}{mJ\,mol^{-1}\,K^{-4}}$
($\Theta_{\mathrm{D}} \approx \SI{358}{K}$). Because the data of
La$_5$Ni$_4$O$_{13}$ scatter considerably below \SI{3}{K}, these values carry
an additional systematic uncertainty; an unweighted fit gives
$\gamma \approx \SI{17}{mJ\,mol^{-1}\,K^{-2}}$.
 
\paragraph{Normal-state background.}
The phonon contribution was described by one Debye and two Einstein terms,
\begin{equation}
  C_{\mathrm{ph}} = C_{\mathrm{D}}(\Theta_{\mathrm{D}}, n_{\mathrm{D}})
                  + C_{\mathrm{E}}(\Theta_{\mathrm{E1}}, n_{\mathrm{E1}})
                  + C_{\mathrm{E}}(\Theta_{\mathrm{E2}}, N - n_{\mathrm{D}} - n_{\mathrm{E1}}),
\end{equation}
with the total number of oscillators fixed to $N = 19$. The parameters were obtained by simultaneously fitting (i) the normal state between 19.5 and \SI{80}{K} with $\gamma_{\mathrm{n}}T + C_{\mathrm{ph}}$, (ii) the data below \SI{4}{K} with $\gamma_0 T + C_{\mathrm{ph}}$, and (iii) the condition of electronic entropy conservation,
\begin{equation}
  \int_0^{T^*} \frac{C_{\mathrm{e}}}{T}\,\mathrm{d}T = \gamma_{\mathrm{n}}T^*,
  \qquad T^* = \SI{20}{K},
\end{equation}
where $C_{\mathrm{e}} = C_p - C_{\mathrm{ph}}$. The fit yields
$\gamma_{\mathrm{n}} = \SI{91}{mJ\,mol^{-1}\,K^{-2}}$,
$\Theta_{\mathrm{D}} = \SI{154}{K}$ ($n_{\mathrm{D}} = 1.6$),
$\Theta_{\mathrm{E1}} = \SI{153}{K}$ ($n_{\mathrm{E1}} = 4.1$) and
$\Theta_{\mathrm{E2}} = \SI{397}{K}$ ($n_{\mathrm{E2}} = 13.3$). The
individual contributions, the fit residuals and the entropy balance are shown in Supplementary Fig.~\ref{fig:SI_HC}b--d; above \SI{80}{K} the model is an extrapolation.
 
\paragraph{Comparison with La$_5$Ni$_4$O$_{13}$.}
The measured $C_p/T$ of La$_5$Ni$_4$O$_{13}$ varies smoothly between 2 and
\SI{30}{K} without any anomaly. Between 10 and \SI{18}{K}, where the
superconducting anomaly of La$_5$Ni$_4$O$_{10}$ appears, it coincides with the
fitted $C_{\mathrm{ph}}/T$ of La$_5$Ni$_4$O$_{10}$ within 3\% (ratio
0.97--1.02; Supplementary Fig.~\ref{fig:SI_HC}e). Above \SI{20}{K} the
ratio decreases gradually, to 0.87 at \SI{30}{K}, which we attribute to the
different lattice dynamics of the two structures. This comparison provides an
independent check of the phonon background in the temperature range of the
superconducting transition.
 
\paragraph{Electronic specific heat.}

We analyzed the electronic specific heat below \SI{22}{K} with the $\alpha$ model, adding a constant term for the non-superconducting part of the sample,

\begin{equation}
  \frac{C_{\mathrm{e}}}{T} = \gamma_0
    + (\gamma_{\mathrm{n}} - \gamma_0)\,g_\alpha(T/T_{\mathrm{c}}).
\end{equation}

Here $g_\alpha$ is the normalized electronic specific heat of a BCS superconductor with a gap $\Delta_0 = \alpha k_{\mathrm{B}}T_{\mathrm{c}}$ ($\alpha = 1.76$ in weak coupling)\cite{padamsee1973quasiparticle}. A single gap with a sharp transition gives
$\gamma_0 = 40.2 \pm \SI{1.0}{mJ\,mol^{-1}\,K^{-2}}$, close to the value measured below \SI{4}{K}, and thus a superconducting volume fraction $f_{\mathrm{SC}} = 1 - \gamma_0/\gamma_{\mathrm{n}} \approx 56\%$. The shape of
the anomaly, however, is not reproduced (reduced $\chi^2 = 6.9$; Supplementary Fig.~\ref{fig:SI_HC}f,g). The measured anomaly is broad and has two maxima, at about 16 and \SI{10}{K}. Adding a Gaussian distribution of $T_{\mathrm{c}}$ ($\sigma = \SI{2.5}{K}$) lowers the reduced $\chi^2$ to 3.7, but the calculated curve still has only one maximum.

Two maxima could come from a single phase with two superconducting gaps, as in MgB$_2$~\cite{bouquet2001specific,bouquet2001phenomenological}, or from two regions with different $T_{\mathrm{c}}$. We tested both, in each case with a
Gaussian distribution of $T_{\mathrm{c}}$. In the two-gap model, both gaps open at the same $T_{\mathrm{c}}$:
\begin{equation}
  \frac{C_{\mathrm{e}}}{T} = \gamma_0 + (\gamma_{\mathrm{n}} - \gamma_0)
  \left[x\,\langle g_{\alpha_1}\rangle + (1 - x)\,\langle g_{\alpha_2}\rangle\right]
\end{equation}
where $x$ is the weight of the smaller gap $\alpha_1$. In the two-component
model, each component has its own $T_{\mathrm{c}}$:
\begin{equation}
  \frac{C_{\mathrm{e}}}{T} = \gamma_0 + (\gamma_{\mathrm{n}} - \gamma_0)
  \left[w\,\langle g_\alpha\rangle_{1} + (1 - w)\,\langle g_\alpha\rangle_{2}\right],
\end{equation}
where $\langle g_\alpha\rangle_{i}$ is averaged over a distribution centred at
$T_{\mathrm{c}i}$ with width $\sigma_i$, and both components share the same
$\alpha$.

In the two-gap model, a direct fit with $\alpha_1$ and $\alpha_2$ left free gives
reduced $\chi^2 = 3.6$, with $\alpha_1 = 0.3$, $\alpha_2 = 3.8$,
$T_{\mathrm{c}} = \SI{15.9}{K}$ and $x = 0.24$ for the smaller gap (dashed green
line in Supplementary Fig.~\ref{fig:SI_HC}f). The smaller gap,
$\Delta_1/k_{\mathrm{B}} = \alpha_1 T_{\mathrm{c}} \approx \SI{5}{K}$, is too
small to explain the maximum near \SI{11}{K}. Further, if we restrict
$\alpha_1$ to $0.6 \le \alpha_1 \le 1.6$ (with $\alpha_2 \ge 1.76$), so that
the smaller gap could account for this maximum, the weight $x$ drops to its
lower bound of 0.02, which is too small to produce a second maximum (dotted
green line in Supplementary Fig.~\ref{fig:SI_HC}f).



The two-component model fits the data better than two-gap model (reduced $\chi^2 = 1.5$, see the orange line in Fig.\ref{fig:SI_HC}f). It gives
$T_{\mathrm{c1}} = \SI{15.9}{K}$ ($\sigma_1 = \SI{2.4}{K}$) and
$T_{\mathrm{c2}} = \SI{10.8}{K}$ ($\sigma_2 = \SI{0.7}{K}$), with 77\% and
23\% of the superconducting fraction, respectively, and
$\gamma_0 = \SI{40}{mJ\,mol^{-1}\,K^{-2}}$, again in line with the
low-temperature value. The fit pushes $\alpha$ to its upper bound, so $\alpha$
itself is not determined, but the transition temperatures, the weights and
$f_{\mathrm{SC}}$ hardly depend on it. Thus, the current data supports more on two-component model.

The specific heat was measured on several crystals at the same time. Since all crystals show the same onset $T_c$ of diamagnetic screening Fig.\ref{squid}, the second component cannot be explained by a few crystals with a lower $T_{\mathrm{c}}$. Resistance is sensitive only to the superconducting path of higher-$T_{\mathrm{c}}$ regions, which explains why transport shows a single transition. From zero-field data alone, we cannot fully rule out multiband superconductivity with weak interband coupling; field-dependent specific heat can help solve this. Because the transitions are broad and only part of the sample is superconducting, we regard the fitted $\alpha$ values as effective parameters and do not analyze the gap size further.

\paragraph{Uncertainty of the superconducting volume fraction.}
Because $\gamma_{\mathrm{n}}$ is obtained from the normal-state fit, we estimated its uncertainty, and that of $f_{\mathrm{SC}}$, by repeating the analysis for different choices of the fitting ranges: lower limits $T_{\mathrm{lo}} = 19.5$, 21 and \SI{23}{K}, upper limits $T_{\mathrm{hi}} = 60$, 80 and \SI{100}{K}, and entropy-balance temperatures $T^* = 20$ and \SI{22}{K}. These choices give $\gamma_{\mathrm{n}}$ between 84 and \SI{92}{mJ\,mol^{-1}\,K^{-2}} and $f_{\mathrm{SC}}$ between 53 and 57\% (Supplementary Fig.~\ref{fig:SI_HC}h). In all cases, more than half of the
sample is superconducting.

\end{document}